\documentclass[trackchanges]{aastex7}

\usepackage{booktabs} 
\usepackage{threeparttable}
\usepackage{multirow} 
\usepackage{array}

\usepackage{xcolor}  
\definecolor{revision}{rgb}{1,0,0} 
\usepackage{ulem}

\usepackage{graphicx}	
\usepackage{amsmath}	
\newcommand\target{{Swift J1727.8--1613}}
\newcommand\hxmt{{\textit{Insight}-HXMT}}

\begin{document}

\title{An {\textit{Insight}-HXMT} view of the evolution of the type-C quasi-periodic oscillation during the flaring state of Swift J1727.8--1613}

\author{Min Wei}
\affiliation{Department of Astronomy, School of Physics and Astronomy,  Yunnan University, Kunming 650091, China}
\email{weimin@stu.ynu.edu.cn}

\author{Xiang Ma} 
\affiliation{Key Laboratory of Particle Astrophysics, Institute of High Energy Physics, Chinese Academy of Sciences, 19B Yuquan Road, Beijing 100049, China}
\email{max@ihep.ac.cn}

\author{Liang Zhang}
\affiliation{Key Laboratory of Particle Astrophysics, Institute of High Energy Physics, Chinese Academy of Sciences, 19B Yuquan Road, Beijing 100049, China}
\email{zhangliang@ihep.ac.cn}

\author{Xiang-Hua Li}
\affiliation{Department of Astronomy, School of Physics and Astronomy,  Yunnan University, Kunming 650091, China}
\email{xhli@ynu.edu.cn}

\author{Ming-Yu Ge}
\affiliation{Key Laboratory of Particle Astrophysics, Institute of High Energy Physics, Chinese Academy of Sciences, 19B Yuquan Road, Beijing 100049, China}
\email{gemy@ihep.ac.cn}

\author{Lian Tao}
\affiliation{Key Laboratory of Particle Astrophysics, Institute of High Energy Physics, Chinese Academy of Sciences, 19B Yuquan Road, Beijing 100049, China}
\email{taolian@ihep.ac.cn}

\author{Jin-Lu Qu}
\affiliation{Key Laboratory of Particle Astrophysics, Institute of High Energy Physics, Chinese Academy of Sciences, 19B Yuquan Road, Beijing 100049, China}
\email{qujl@ihep.ac.cn}

\author{Shuang-Nan Zhang}
\affiliation{Key Laboratory of Particle Astrophysics, Institute of High Energy Physics, Chinese Academy of Sciences, 19B Yuquan Road, Beijing 100049, China}
\email{zhangsn@ihep.ac.cn}

\author{Shu Zhang}
\affiliation{Key Laboratory of Particle Astrophysics, Institute of High Energy Physics, Chinese Academy of Sciences, 19B Yuquan Road, Beijing 100049, China}
\email{szhang@ihep.ac.cn}

\author{Li-Ming Song}
\affiliation{Key Laboratory of Particle Astrophysics, Institute of High Energy Physics, Chinese Academy of Sciences, 19B Yuquan Road, Beijing 100049, China}
\email{songlm@mail.ihep.ac.cn}

\author{Rui-Can Ma}
\affiliation{Key Laboratory of Particle Astrophysics, Institute of High Energy Physics, Chinese Academy of Sciences, 19B Yuquan Road, Beijing 100049, China}
\email{maruican@mail.ihep.ac.cn}

\author{Zi-Xu Yang}
\affiliation{School of Physics and Optoelectronic Engineering, Shandong University of Technology, Zibo 255000, China}
\email{yangzx@sdut.edu.cn}

\author{Yue Huang}
\affiliation{Key Laboratory of Particle Astrophysics, Institute of High Energy Physics, Chinese Academy of Sciences, 19B Yuquan Road, Beijing 100049, China}
\email{huangyue@mail.ihep.ac.cn}

\author{Pan-Ping Li}
\affiliation{Key Laboratory of Particle Astrophysics, Institute of High Energy Physics, Chinese Academy of Sciences, 19B Yuquan Road, Beijing 100049, China}
\affiliation{University of Chinese Academy of Sciences, Chinese Academy of Sciences, Beijing 100049, China}
\email{lipp@mail.ihep.ac.cn}

\author{Jia-Ying Cao}
\affiliation{Key Laboratory of Particle Astrophysics, Institute of High Energy Physics, Chinese Academy of Sciences, 19B Yuquan Road, Beijing 100049, China}
\affiliation{University of Chinese Academy of Sciences, Chinese Academy of Sciences, Beijing 100049, China}
\email{caojy@ihep.ac.cn}

\author{Shu-Jie Zhao}
\affiliation{Key Laboratory of Particle Astrophysics, Institute of High Energy Physics, Chinese Academy of Sciences, 19B Yuquan Road, Beijing 100049, China}
\affiliation{University of Chinese Academy of Sciences, Chinese Academy of Sciences, Beijing 100049, China}
\email{zhaoshujie@mail.ihep.ac.cn}

\author{Qing-Chang Zhao}
\affiliation{Key Laboratory of Particle Astrophysics, Institute of High Energy Physics, Chinese Academy of Sciences, 19B Yuquan Road, Beijing 100049, China}
\affiliation{University of Chinese Academy of Sciences, Chinese Academy of Sciences, Beijing 100049, China}
\email{zhaoqc@mail.ihep.ac.cn}

\author{Yun-Xiang Xiao}
\affiliation{Key Laboratory of Particle Astrophysics, Institute of High Energy Physics, Chinese Academy of Sciences, 19B Yuquan Road, Beijing 100049, China}
\affiliation{University of Chinese Academy of Sciences, Chinese Academy of Sciences, Beijing 100049, China}
\email{xiaoyunxiang@mail.ihep.ac.cn}

\author{Guo-Li Huang}
\affiliation{Key Laboratory of Particle Astrophysics, Institute of High Energy Physics, Chinese Academy of Sciences, 19B Yuquan Road, Beijing 100049, China}
\affiliation{University of Chinese Academy of Sciences, Chinese Academy of Sciences, Beijing 100049, China}
\email{huanggl@ihep.ac.cn}

\correspondingauthor{Xiang Ma}
\email[show]{max@ihep.ac.cn}

\correspondingauthor{Liang Zhang}
\email[show]{zhangliang@ihep.ac.cn}

\correspondingauthor{Xiang-Hua Li}
\email[show]{xhli@ynu.edu.cn}


\begin{abstract}

We present a detailed analysis of the evolution of type-C quasi-periodic oscillations (QPOs) observed during the flaring state of the recently discovered black hole X-ray binary \target, utilizing data from \hxmt.
By examining the relation between the QPO fractional rms amplitude and QPO frequency across various energy bands, we discover that the behavior significantly differs between these energy bands.
Below 10 keV, the QPO fractional rms generally decreases with increasing QPO frequency, whereas above 10 keV, the QPO fractional rms remains relatively stable with frequency.
Additionally, we report, for the first time, the detection of a common break at around 4 Hz in the relation between QPO fractional rms and frequency in both the 2--4 keV and 50--100 keV energy bands.  
We also find that the evolution of all the spectral parameters alters its behavior at around 4 Hz, with the changes in all parameters becoming flatter. This suggests a significant change in the geometry of the accretion flow.
We attribute the observed break to the overall changes in the spectrum.

\end{abstract}



\section{Introduction}

Black hole transients (BHTs) are binary systems consisting of a stellar-mass black hole accreting matter from \textbf{a low-mass ($<$\ 1 $\mathrm{M_\odot}$)} companion star via Roche lobe overflow. These systems are typically transient sources, remaining in a quiescent state for the majority of their lifespan, with occasional outbursts during which their X-ray luminosity increases by several orders of magnitude (see \citealt{Belloni2016Transient_BHBs} for a recent review). 


BHTs in outbursts are characterized by the rapid evolution of their X-ray spectral and timing properties, which can be categorized into distinct phases: Quiescent State (QS), Low/Hard State (LHS), Hard-intermediate State (HIMS), Soft-intermediate state (SIMS), and High/Soft State (HSS). Additionally, some sources may also display a Very High State (VHS) \citep{Belloni2005Theevolution, Homan2005,Belloni2016Transient_BHBs}. The progression through different spectral states is closely linked to variations in the mass accretion rate, offering valuable insights into the changing dynamics of the accretion flow geometry.
During most outbursts, BHTs evolve from the LHS through the HIMS and SIMS before reaching the HSS during the outburst rise. In the decay phase, they return to the LHS in reverse order. This evolutionary path displays a counterclockwise `q'-shaped pattern in the hardness-intensity diagram (HID).
In the LHS, the X-ray spectrum is dominated by a hard power-law component with a high-energy cutoff \textbf{that can be up to $\sim$100 keV}. In contrast, the spectrum of the HSS is dominated by a soft component associated with an optically thick, geometrically thin accretion disk \textbf{when the Eddington ratio is below 30\%}. Meanwhile, the HIMS and SIMS spectra show contributions from both the hard and soft components.

Low-frequency quasi-periodic oscillations (LFQPOs) with centroid frequencies ranging from a few mHz to 30 Hz are common features in the power density spectra (PDS) of BHTs (see \citealt{Ingram2020BHXBreview} for a recent review). Based on differences in the shape of the PDS, LFQPOs in BHTs can be classified into three distinct types: type-A, type-B, and type-C \citep{Remillard2002Characterizing_QPO,Casella2005ABC_LF}. 
Type-A QPOs are typically observed in the HSS and are characterized by weak and broad peaks around 6--8 Hz \citep{Zhang2023type-AQPO}. Type-B QPOs, associated with the SIMS, appear as relatively strong and narrow peaks with a centroid frequency of 4--6 Hz \citep{Stevens2016,Gao2017,Homan2020,LiuHX2022type-BQPO,Zhang2023}. Both type-A and type-B QPOs are accompanied by weak broadband noise. Type-C QPOs, the most common type of LFQPOs, are most frequently detected in the LHS and HIMS. These QPOs are characterized by high-amplitude, narrow peaks with variable frequencies ranging from 0.1 to 30 Hz, along with a strong broadband noise continuum. Multiple harmonic components are occasionally observed in the PDS \citep{Motta2011,Motta2015,Zhang2020,Ma2021NatAs...5...94M,Alabarta2022}. 

Over the last few decades, the analysis of X-ray spectro-timing correlations has been shown to play a significant role in indirectly mapping the accretion region. For instance, the observations of MAXI J1535-571 suggest the variation between spectral parameters and QPO frequency may be associated with changes in the corona or jet \citep{Rawat2023maxi1535,Zhangyx20221535}. Therefore, the study of relationship between timing or energy spectral parameters and QPO frequency can provide critical insights into the physical origin of QPOs and the geometric evolution of the disk-corona system.

\target\ was first detected on 2023 August 24 by {\it Swift/BAT} \citep{Page2023_1727}. Subsequent multiwavelength observations have identified the source as a bright BHT with a peak flux of exceeding 7 Crab in the MAXI 2--20 keV band \citep{Baglio2023optical, Miller-Jones2023,Negoro2023maxi_bht,Sunyaev2023}. The distance to the source has been estimated at $5.5_{-1.1}^{+1.4}$~kpc (\citealt{Burridge2025distance1727}). 
An \textbf{inner disk} inclination angle of $\sim40^{\circ}$ was derived from the spectral modeling of the reflection component (\citealt{Peng20241727hardtail}). 
The evolution of the spectral-timing properties of \target\ in the LHS and HIMS has been studied by \citet{yu2024timing}. The source exhibited a rapid increase in flux, followed by a gradual decay and several flares \textbf{in the HIMS}. Strong type-C QPOs with frequencies ranging from $\sim$0.1 to $\sim$8 Hz have been detected \citep{Chatterjee2024,Shui2024,yu2024timing,Zhu2024Energydependence}. Additionally, the X-ray spectrum revealed a high-energy hard tail extending beyond 100 keV in addition to the standard thermal Comptonization component  (\citealt{Mereminskiy2024..250keV, Peng20241727hardtail, 
Yangzx2024ApJ1727}). 

In this work, we investigate the characteristics of the type-C QPOs observed in \target\ during the flaring period of the outburst, using data from the Hard X-ray Modulation Telescope (\hxmt, \citealt{Zhang2020HXMT}). 
We describe the observations and data reduction procedures in Section 2. We present the results of our study in Section 3 and discuss the implications for the observed evolution in Section 4.

\section{Observations and Data Reduction}

\hxmt\ is China's first X-ray astronomy telescope launched on 2017 June 15 \citep{Zhang2020HXMT}. It is equipped with three main instruments: the low-energy detector (LE, 1--12 keV, \citealt{Chen2020SCPMA..6349505CLE}), the medium-energy detector (ME, 8--30 keV, \citealt{Cao2020SCPMA..6349504CME}), and the high-energy detector (HE, 20--250 keV, \citealt{Liu2020SCPMA..6349503LHE}). Additionally, \hxmt\ includes a Space Environment Monitor that tracks charged particles and provides background data for estimation. 

\begin{figure*}       
	\centering
    \includegraphics[width=1\columnwidth]{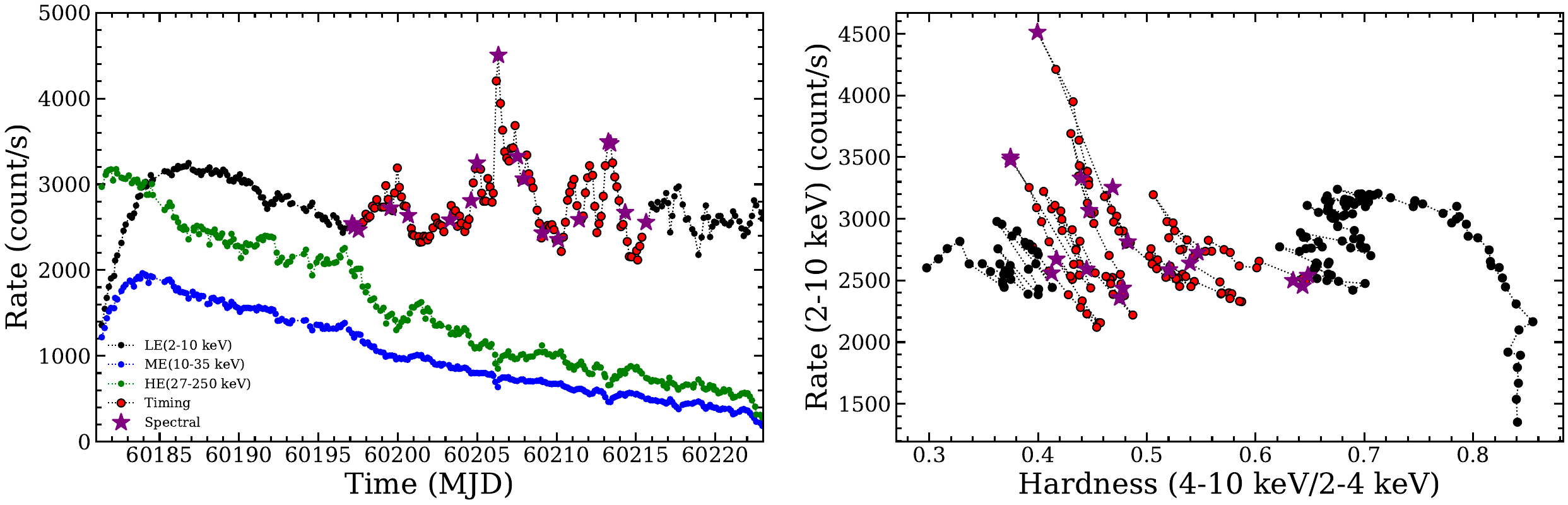}  
	\caption{
    {\it Left}: the LE (2--10 keV), ME (10--35 keV), and HE (27--250 keV) light curves of \target\ observed by \hxmt. 
{\it Right}: the hardness-intensity diagram of \target\ observed by \hxmt. The hardness is defined as the ratio between the 2--4 keV and 4--10 keV count rates. 
Each data point corresponds to an exposure ID.
The red points denote the data used for our timing analysis, while the purple stars mark the data selected for spectral fitting.
    }   
	\label{FIG:lchid}
\end{figure*}

\hxmt\ has extensively observed the outburst of \target\ from 2023 August 25 to 2023 October 4.
In Figure \ref{FIG:lchid}, we show the light curves of the three instruments of \hxmt\ and the HID for the outburst of \target. We refer to \cite{yu2024timing} for details on the spectral-timing evolution.
Between MJD 60197 and MJD 60215, several flares can be seen in the LE light curve, each lasting several days. During these flares, the ME and HE light curves display corresponding declines. 
Hereafter, we will focus our analysis on the observations taken during the flaring period.  These observations correspond to ObsIDs P0614338011 -- P0614338029.
Note that each \hxmt\ observation (ObsID) is split into multiple segments (named ExpID).

We processed the data using the \hxmt\ software package {\it hxmtsoft} 2.06 and applied the following filtering criteria: (1) pointing offset angle less than $\sim0.04^{\circ}$; (2) Earth elevation angle larger than $\sim10^{\circ}$; (3) the value of the geomagnetic cutoff rigidity larger than 8\,GeV; (4) at least 300\,s before and after the passage of the South Atlantic Anomaly. To avoid potential contamination from bright earth and nearby sources, we only use data from the small field of view detectors. 
\textbf{All the errors quoted in this work are at the 90\% confidence level.}

\section{Analysis and Results}
\subsection{Timing analysis}

\subsubsection{Dynamical and average power spectra}

For our timing analysis, we created PDS in different energy bands for each ExpID. We used a 64\,s long interval and a 1/128\,s time resolution. The resulting PDS were then applied to Miyamoto normalization \citep{Miyamoto1991}.


\begin{figure}      
	\centering
    \includegraphics[width=0.6\columnwidth]{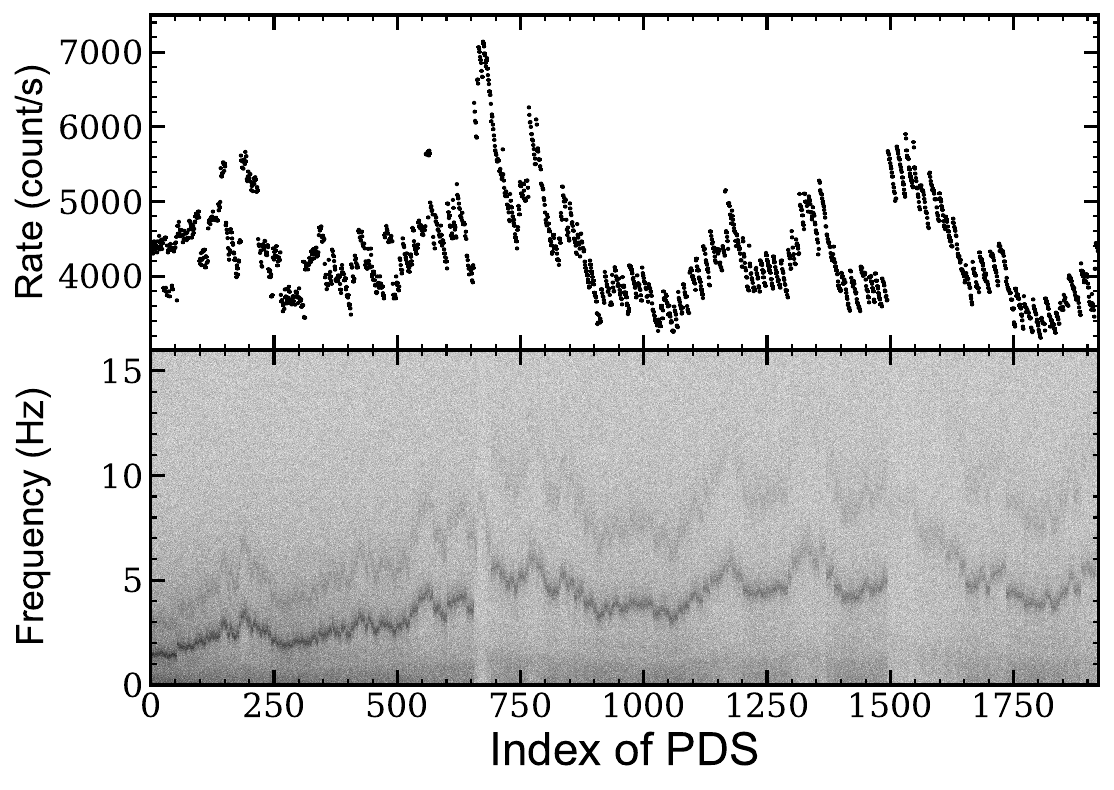}  
	\caption{
 \hxmt\ LE 2--10 keV light curve and corresponding dynamical power spectrum with a time resolution of 64
  s for the flaring period of \target. All time gaps were removed and the $x-$axis represents the index of the PDS.
    } 
	\label{FIG:DPS}
\end{figure}

\begin{figure}      
	\centering
    \includegraphics[width=0.50\columnwidth]{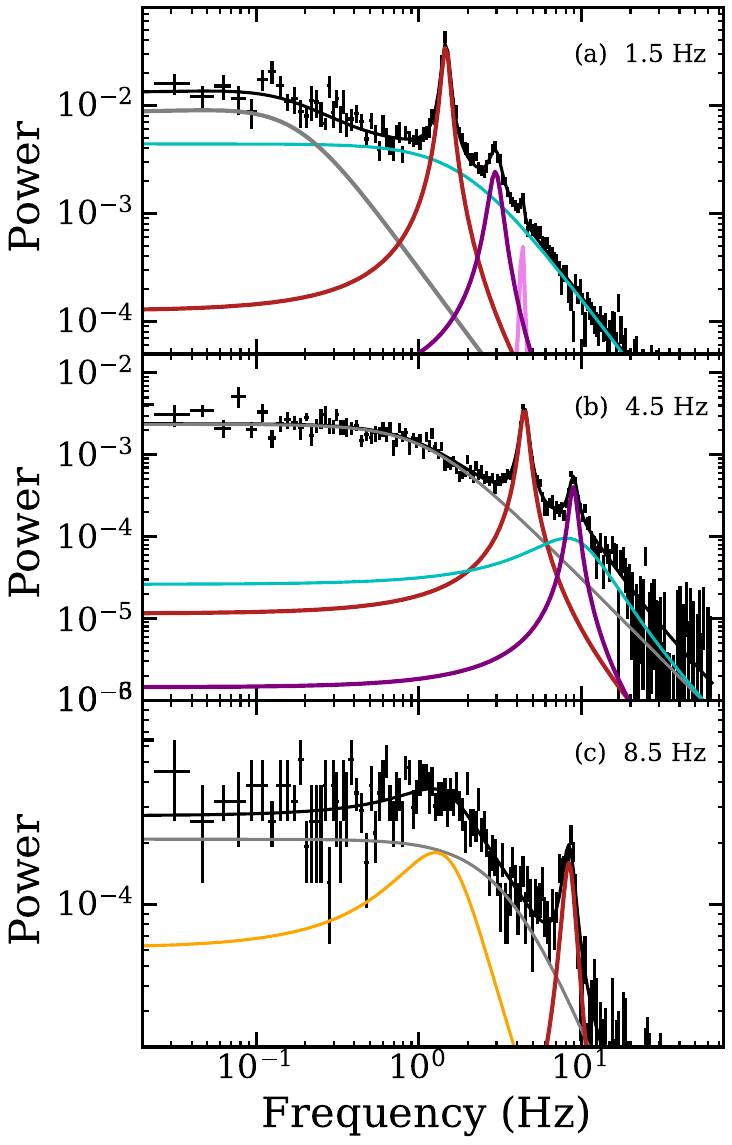}  
	\caption{
    Representative power density spectra for ExpIDs with different QPO frequencies. The PDS were calculated in the 2--10 keV energy band and fitted with a model composed of multiple Lorenzian functions
    The PDS were rms-normalized and the contribution due to Poisson noise was subtracted. The data we used are ExpIDs P061433801101, P061433802503, and P061433802702.
    }   
	\label{FIG:pds}
\end{figure}

In Figure~\ref{FIG:DPS}, we show the LE 2--10 keV light curve and corresponding dynamical PDS with a time resolution of 64 s for the flaring period of \target. All time gaps were removed and the $x$-axis marks the index of each 64-s PDS. 
A significant QPO with a variable frequency is evident in the dynamical PDS. Most of the time, the QPO is accompanied by a second harmonic.


In Figure~\ref{FIG:pds}, we show three representative PDS exhibiting QPOs with different frequencies. Following \cite{Belloni2002}, we fitted the QPO and its harmonics with several narrow Lorentzian functions. \textbf{Based on the fitting, we excluded from our analysis non-significantly detected peaks with a significance\footnote{The significance of the QPO is given as the ratio of the integral of the power of the Lorentzian used to fit the QPO divided by its 1$\sigma$ error.} of less than 3$\sigma$. The QPOs we detected have frequencies in the range of $\sim1.5-8.5$ Hz, with a $Q$ factor spanning from 3 to 12. The fractional rms of the QPO was measured from the fits using the formula rms = $\sqrt{P} \times (S + B)/S$ \citep{Bu2015ApJ...799....2B}, where $S$ and $B$ represent the source and background count rates, respectively, and $P$ is the integrated power from the Miyamoto normalization. 
The low-frequency noise component is well described by a zero-centered Lorentzian, while the high-frequency part can be fitted with one or two broad Lorentzian functions. The level of Poisson noise was modeled as a constant and subsequently subtracted from the PDS. }

\begin{figure}      
	\centering
    \includegraphics[width=0.6\columnwidth]{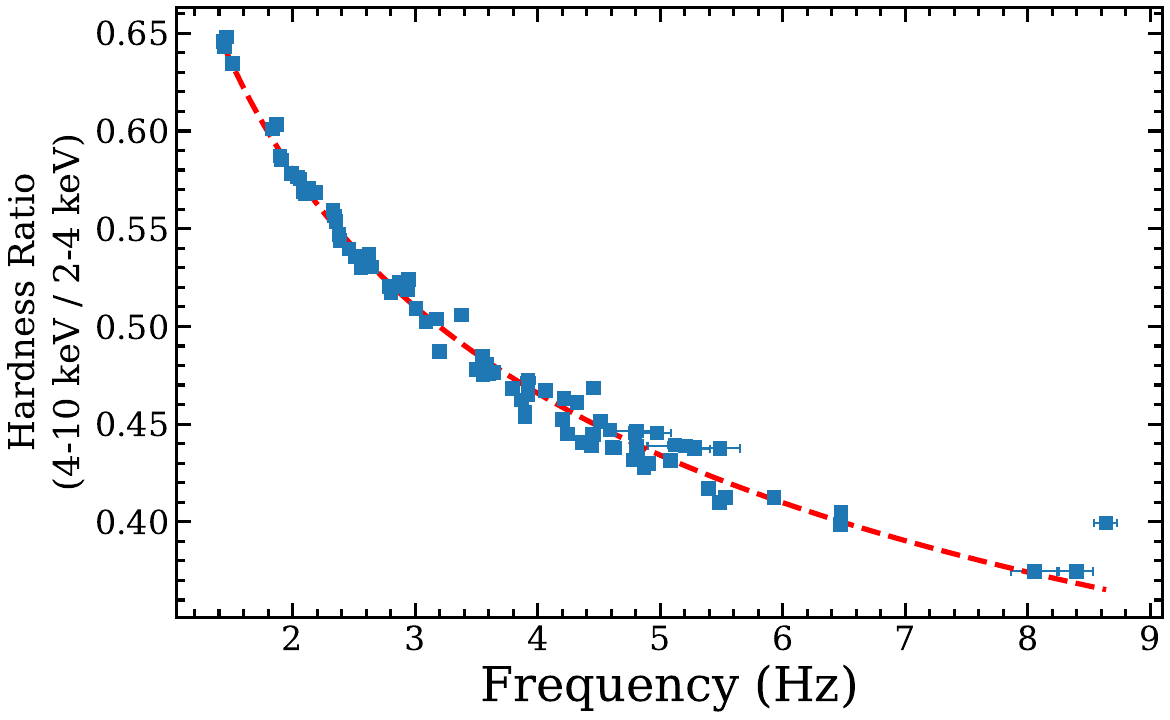}  
	\caption{Hardness ratio versus QPO frequency for the observations in the flaring state of \target. The hardness ratio is defined as the ratio of the count rate between the 4--10 keV and 2--4 keV. The red dotted line is the best-fitting curve using a power-law function. 
}   
	\label{FIG:hardness-frequency}
\end{figure}

\textbf{
In Figure \ref{FIG:hardness-frequency}, we show the hardness ratio (4--10 keV/2--4 keV) versus the QPO frequency for the observations in the flaring state. It is apparent that the QPO frequency is strongly anti-correlated with the hardness ratio.
This relation can be well described by a power-law function of the form $\mathrm{HR} = a* \nu_{\mathrm{QPO}}^{n}$, with the best-fitting parameters $n = -0.329\pm0.005$ and $a=0.730\pm 0.005$. 
%
%
}

\subsubsection{QPO fractional rms versus energy }

\begin{figure}      
	\centering
    \includegraphics[width=0.6\columnwidth]{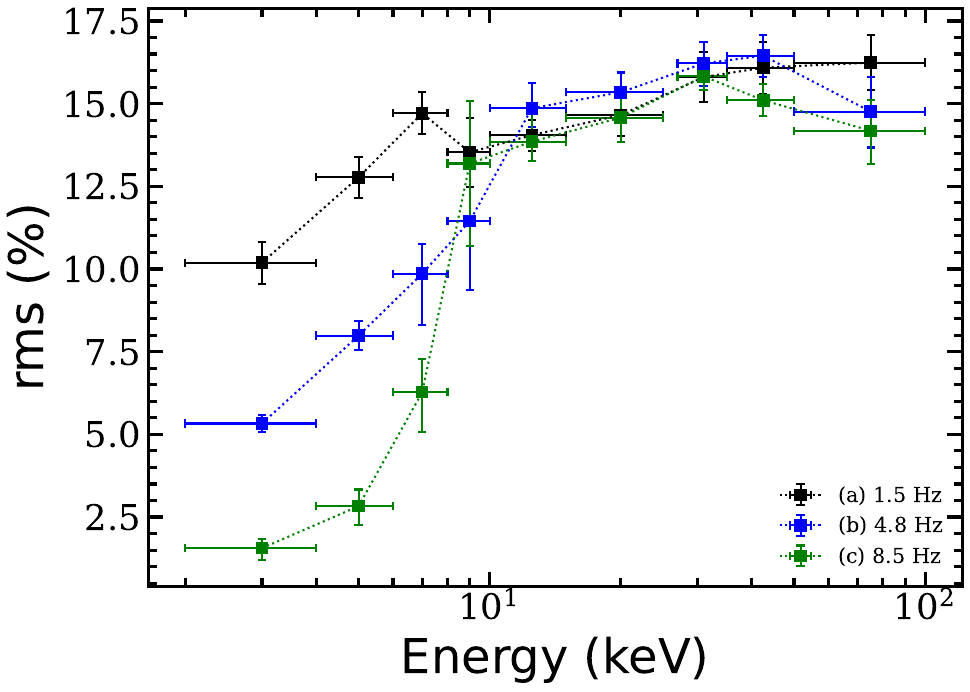}  
	\caption{QPO fractional rms as a function of photon energy for ExpIDs with different QPO frequencies. Black, blue, green points correspond to the ExpIDs P061433801101, P061433802405, and P061433802702, respectively.}   
	\label{FIG:rms-energy}
\end{figure}

We also studied the energy dependence of the QPO fractional rms. In Figure~\ref{FIG:rms-energy}, we show the fractional rms as a function of photon energy for QPOs with different frequencies.
Below approximately 10\,keV, the QPO fractional rms gradually increases with increasing energy while decreasing with increasing frequency.
In the energy range of $\sim$10--30 keV, the increasing trend of the QPO rms with energy flattens in all cases.
Above $\sim$30 keV, QPOs with lower and higher frequencies exhibit different behaviors. The fraction rms of the QPO with a low frequency (1.5-Hz QPO) is significantly higher than that observed below 30 keV, as detailed in the study by \citet{Yangzx2024ApJ1727}. In contrast, the fractional rms of the QPOs with high frequencies (4.8-Hz and 8.5-Hz QPOs) shows a decreasing trend with increasing energy.

\subsubsection{QPO fractional rms versus Frequency}

\begin{figure*}      
	\centering
    \includegraphics[width=1\columnwidth]{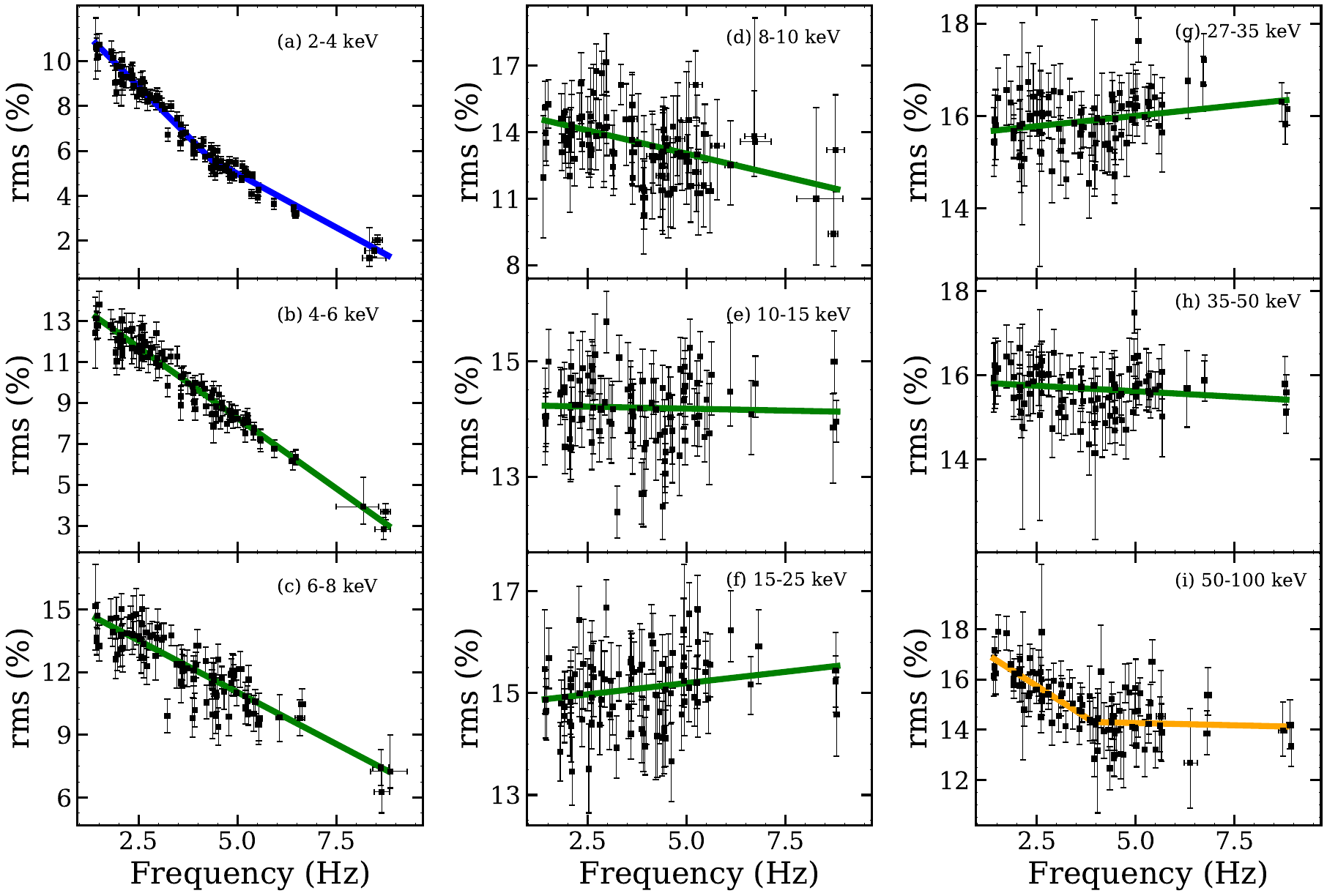}  
	\caption{
    QPO fractional rms amplitude as a function of QPO frequency for different energy bands. The solid line represents the best-fitting model for the data. See the text for details.
    }   
	\label{FIG:rms-frequency}
\end{figure*}

\begin{table}
	\centering
	\caption{
    The Pearson correlation coefficient and the null hypothesis probability for the relation between the QPO fractional rms and centriod frequency across various energy bands shown in  Fig.~\ref{FIG:rms-frequency}.
    }
	\label{tab:1}
	\begin{tabular}{ccc} 
		\hline
		Energy band  &   $r$   &     $p$   \\
		\hline
        2--4 keV & $-0.974$ & $2.04\times 10^{-59}$ \\
	  4--6 keV & $-0.970$ & $1.01\times 10^{-56}$ \\
	  6--8 keV & $-0.875$ & $1.02\times 10^{-29}$ \\
        8--10 keV & $-0.442$ & $1.15\times 10^{-5}$ \\
        10--15 keV & $-0.029$ &   $0.784$ \\
        15--25 keV & $0.209$  &   $0.047$  \\
        27--35 keV & $0.258$  &   $0.014$ \\
        35--50 keV & $-0.124$  &   $0.246$  \\
        50--100 keV & $-0.624$  & $4.91\times 10^{-11}$ \\
		\hline
	\end{tabular}
\end{table}

\begin{figure}      
	\centering
    \includegraphics[width=0.6\columnwidth]{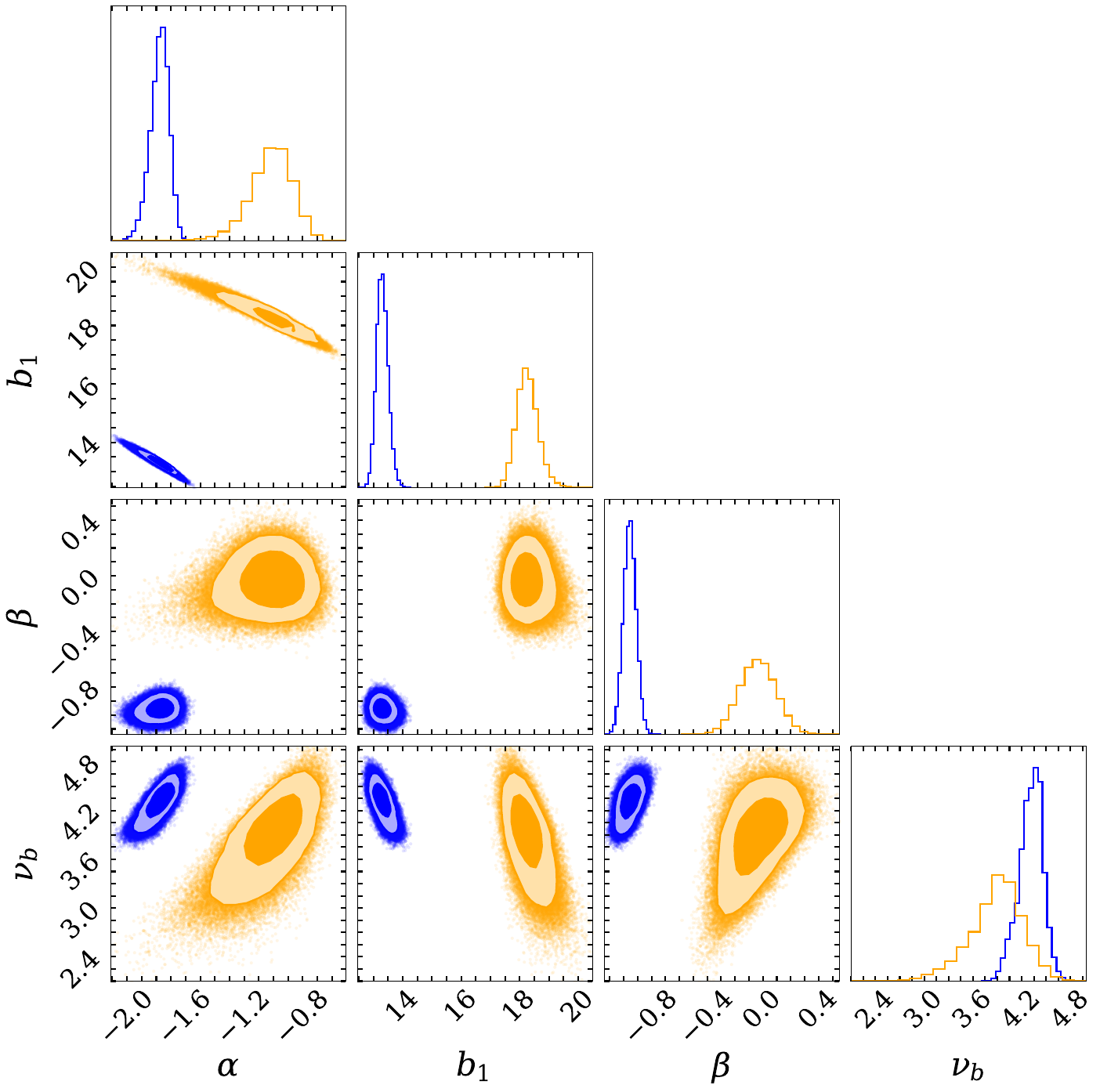}  
	\caption{
    Distributions of the parameters obtained by fitting the relation between the QPO fractional rms and its frequency in the 2--4 keV and 50--100 keV energy bands with the broken-line model using the MCMC method.
    }   
	\label{FIG:mcmc-rms}
\end{figure}

In Figure~\ref{FIG:rms-frequency}, we show the evolution of the QPO fractional rms in relation to its centroid frequency across various energy ranges.
To achieve this, we fitted the PDS in the LE (2--4, 4--6, 6--8, 8--10\,keV), ME (10--15, 15--25\,keV), and HE (27--35, 35--50, 50--100\, keV) bands for each ExpID. The Pearson correlation coefficient and the null hypothesis probability for this relation across the various energy bands are listed in Table~\ref{tab:1}. 
A clear anti-correlation is observed in the LE 2--4 keV, 4--6 keV, and 6--8 keV bands. However, this correlation becomes much weaker in the LE 8--10 keV band. In the ME 10--15 keV, 15--25 keV, and HE \textbf{27--35 keV,} 35--50 keV bands, the anti-correlation between the QPO fractional rms and its frequency disappears, with the QPO rms remaining nearly constant across different QPO frequencies.
In the HE 50--100 keV band, an anti-correlation is apparent again.

From this figure, it is also evident that the anti-correlation between the QPO fractional rms and its frequency shows a distinct break around 4 Hz in the LE 2--4 keV and HE 50--100 keV bands. However, this break is not significant in the other energy bands. To confirm this,  we fitted the relation using both a broken-line and a straight-line model for each energy band. We found that only in the LE 2--4 keV and HE 50--100 keV bands does the broken-line model significantly outperform the straight-line model at a significance of $\ge3\sigma$.
Using the Markov Chain Monte Carlo (MCMC) method, \textbf{we modeled the relation between the QPO fractional rms ($R$) and its frequency ($\nu$) in the LE 2--4 keV and HE 50-100 keV bands with a broken-line function that remains continuous at the break frequency $\nu_b$:
\begin{equation}
R(\nu) = 
\begin{cases}
\alpha \cdot \nu + b_1, & \text{if } \nu < \nu_b \\
\beta \cdot \nu + b_2, & \text{if } \nu \ge \nu_b
\end{cases}
\end{equation}
where $b_2 = \alpha \cdot \nu_b + b_1 - \beta \cdot \nu_b$ to ensure continuity at $\nu = \nu_b$. The parameters $\alpha$ and $\beta$ represent the slopes below and above the break, respectively.}
%
%
Distributions of the parameters are shown in Figure~\ref{FIG:mcmc-rms}. For the 2--4 keV band, we obtained a slope below the break as \textbf{$\alpha_{2-4} = -1.77_{-0.07}^{+0.06}$, the slope above the break as $\beta_{2-4} = -0.96_{-0.04}^{+0.05}$, and the break frequency as $\nu_{b,2-4} = 4.32_{-0.16}^{+0.13}$ Hz. The corresponding values for the 50--100 keV band are $\alpha_{50-100} = -1.00_{-0.15}^{+0.13}$, $\beta_{50-100} = -0.04\pm{0.13}$, and $\nu_{b,50-100} = 3.93_{-0.35}^{+0.27}$ Hz}.
Notably, the break frequencies in the 2--4 keV and 50--100 keV bands are consistent within errors.

\subsection{Spectral analysis}

Given that we detected a common break in the correlation between the QPO fractional rms and its frequency in both the 2--4 keV and 50--100 keV bands, it is essential to check whether the spectral parameters vary around the break frequency. 
Therefore, we performed spectral fitting for 18 ExpIDs that cover the entire QPO frequency range. 
\textbf{We selected these observations based on roughly uniform sampling, choosing data approximately every 1 Hz near the break point. To minimize the influence of any single ExpID on the results, three ExpIDs were included for each frequency range.}
The spectra were rebinned to ensure a minimum of 200 counts per bin for LE and at least 50 counts per bin for both ME and HE.
The energy bands adopted for spectral analysis are LE 2--10 keV, ME 10--30 keV, and HE 28--120 keV.

Following \citet{Yangzx2024ApJ1727}, we fitted the spectra with the model: \texttt{constant * tbabs * (diskbb + relxill + cutoffpl)}.
The \texttt{constant} reflects the relative calibration between different instruments and was fixed at 1 for LE.
The \texttt{tbabs} component is used to account for interstellar absorption along the line of sight, using abundance values from \citet{Wilms2000ApJ...542..914Wtbabs}. 
The \texttt{diskbb} represents a multicolor blackbody component from the accretion disk (\citealt{Mitsuda1984PASJ...36..741M..diskbb}). 
The \texttt{relxill} component models the standard thermal Comptonization component with reflection (\citealt{Dauser2016AN....337..362D..relxill}). 
Finally, the \texttt{cutoffpl} is used to describe the additional high-energy component observed in the spectra. 

\begin{figure}       
	\centering
    \includegraphics[width=0.6\columnwidth]{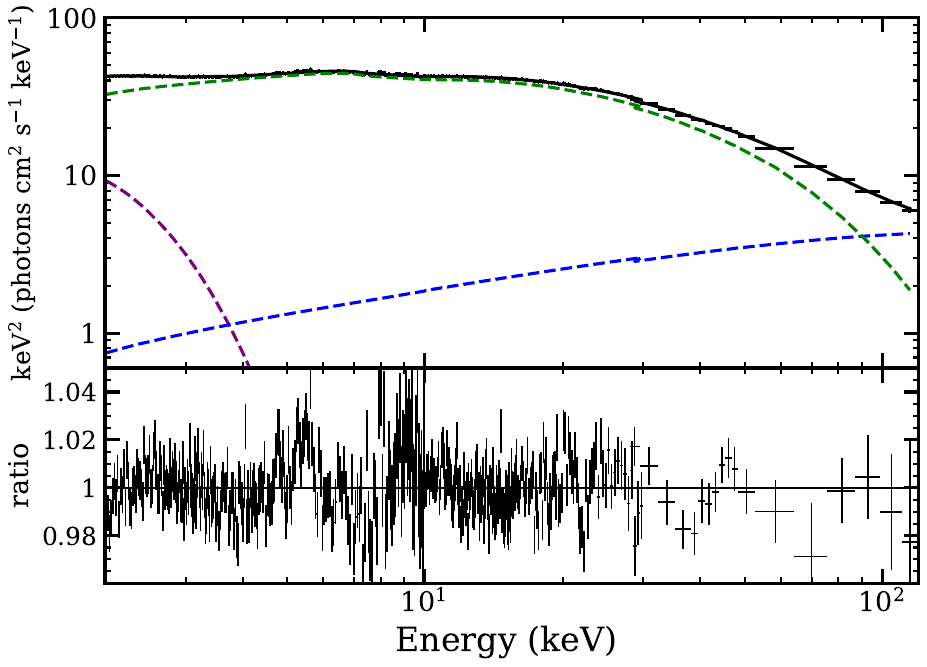}  
	\caption{
    A representative unfolded spectrum and residuals for ExpID P061433801101. The spectrum was fitted using the model \texttt{constant * Tbabs * (diskbb + relxill + cutoffpl)}.
    }   
	\label{FIG:spec fitting}
\end{figure}

\begin{figure}       
	\centering
    \includegraphics[width=0.9\columnwidth]{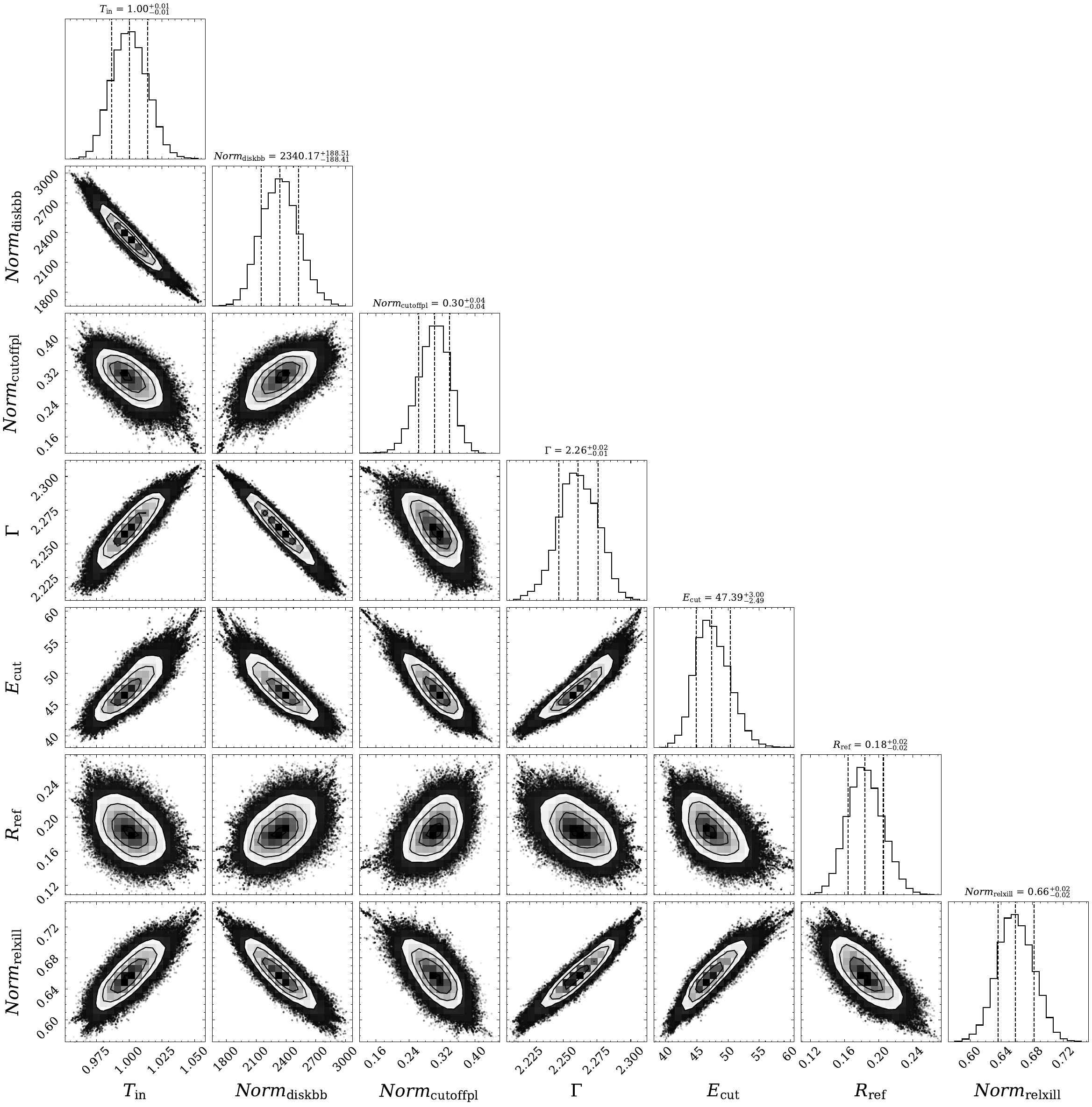} 
	\caption{ 
    Corner plot of the posterior distributions for all free parameters derived from the MCMC spectral fitting (ExpID P061433801807).
    }   
	\label{FIG:corner_plot}
\end{figure}

Focusing on the evolution of spectral parameters during the flaring state, we fixed several parameters that are unlikely to change over timescales of a few days throughout the fitting process.
The hydrogen column density, $N_{\rm H}$, was fixed at $0.3 \times 10^{22}~{\rm cm}^{-2}$ following \citet{Liu2024hardtail}.
In the \texttt{relxill} model, the black hole spin and the inclination angle were fixed at the values obtained by \citet{Peng20241727hardtail}. Additionally, the iron abundance, $A_{\text{Fe}}$, is fixed at \textbf{1.17} based on the value reported by \citet{Yangzx2024ApJ1727}. 
The inner disk radius, $R_{\rm in}$, was allowed to vary initially. However, for ExpIDs with QPO frequencies greater than 1.5 Hz, the best-fitting values of  $R_{\rm in}$ consistently reach the ISCO. Therefore, during these ExpIDs, we fixed $R_{\rm in}$ at the ISCO.
Similarly, the ionization parameter, ${\rm log} \xi$, of these observations was fixed at 3.7, which is the value derived from the initial three observations. \textbf{Fixing the ionization parameter at either 3.5 or 4.0 does not affect the evolution of the main spectral parameters.}
%
The additional hard component (\texttt{cutoffpl}) is very weak during the flaring state, and its parameters are poorly constrained. However, the F-test indicates that this component is significantly needed in most cases. Therefore, we fixed the power-law photon index, $\Gamma$, at 1.5 and the high-energy cutoff, $E_{\rm cut}$, at 300 keV, and allowing only the normalization to vary.
The best-fitting spectral parameters are presented in Table \ref{tab:2}, and a representative unfolded spectrum along with its best-fitting model is displayed in Figure~\ref{FIG:spec fitting}.
\textbf{In Figure~\ref{FIG:corner_plot}, we show the distributions of the main spectral parameters obtained from the fit to the spectrum of ExpID P061433801807.}

\begin{figure}       
	\centering
    \includegraphics[width=0.5\columnwidth]{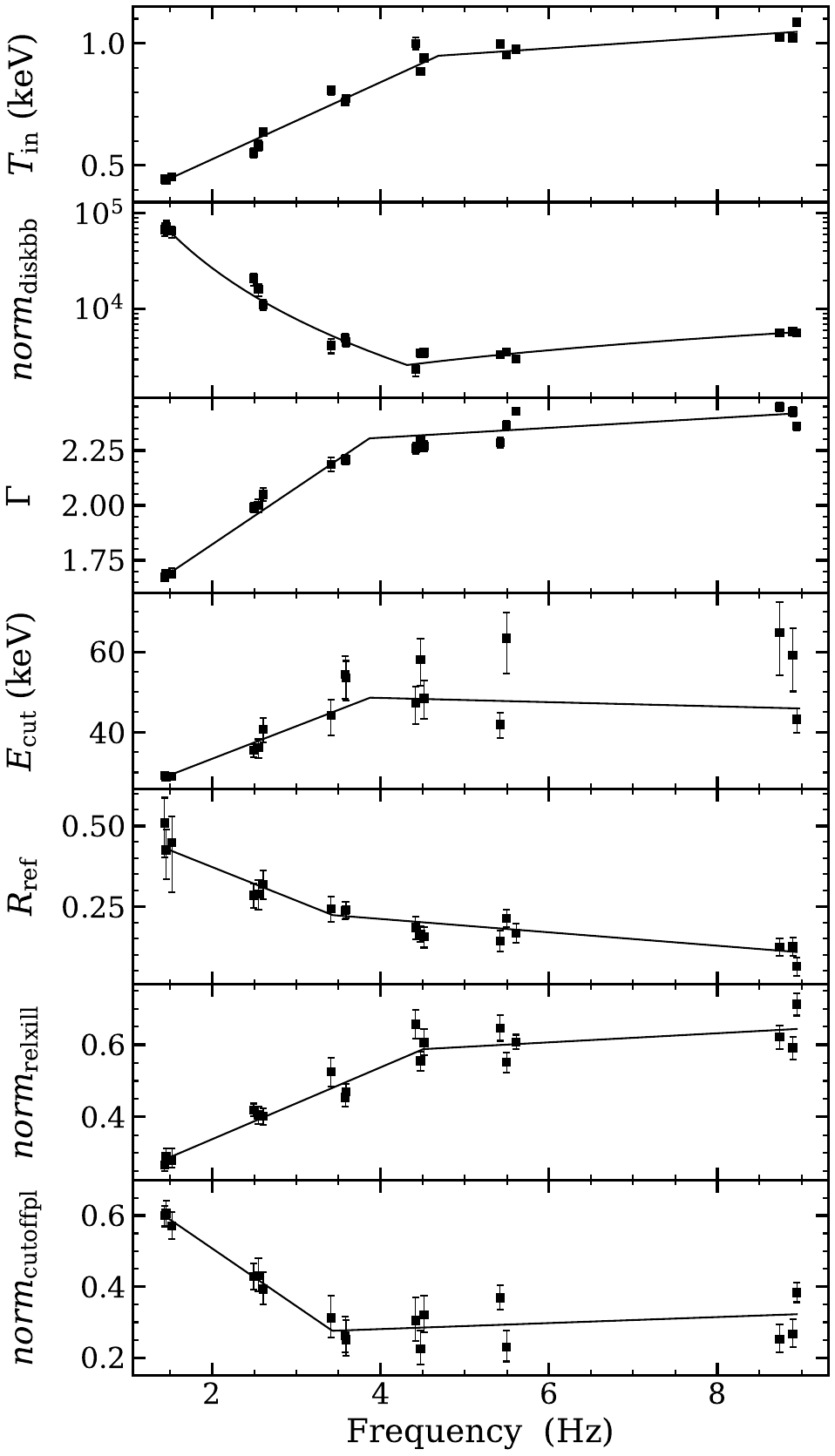} 
	\caption{ 
    Evolution of the main spectral parameters with QPO frequency. The spectra were fitted using the model \texttt{constant * Tbabs * (diskbb + relxill + cutoffpl)}. From top to bottom: inner disk temperature, \texttt{diskbb} normalization, power-law photon index, high-energy cutoff, reflection fraction, \texttt{relxill} normalization, and \texttt{cutoffpl} normalization. The solid line represents the best-fitting broken (or broken power-law) model for the data.
    }   
	\label{FIG:specpara}
\end{figure}

\begin{figure}       
	\centering
    \includegraphics[width=0.55\textwidth]{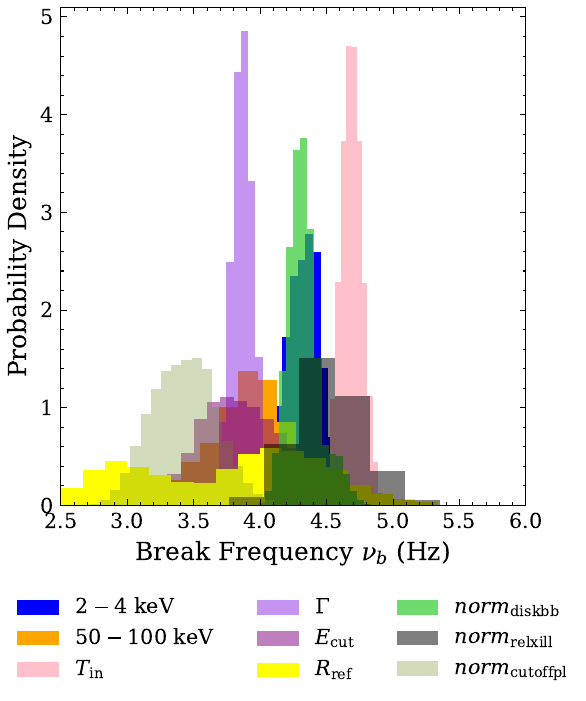}  
	\caption{
    Distribution of the best-fitting values for the break frequency in the relations between different spectral parameters and QPO frequency, as derived from MCMC analysis.
    }   
	\label{FIG:turning points}
\end{figure}

In Figure~\ref{FIG:specpara}, we show the evolution of the main spectral parameters as a function of QPO frequency. 
It is evident that all the parameters exhibit a break at approximately 4 Hz. However, it is noteworthy that their break frequencies are not entirely consistent.
The parameters $T_{\rm in}$, $\Gamma$, $E_{\rm cut}$, and the normalization of the \texttt{relxill} component initially increase with the QPO frequency before plateauing. In contrast, the normalization of the \texttt{diskbb} component, $R_{\rm ref}$, and the normlization of the additional hard component exhibit an initial decrease with QPO frequency, followed by a flattening trend.
We fitted the relations between different spectral parameters and QPO frequency using either a broken-line or broken power-law model.
In Figure~\ref{FIG:turning points}, we show the probability density distributions of the break frequencies, as derived from MCMC analysis.
Our findings reveal that the break frequencies for the relations between $\Gamma$, $E_{\rm cut}$, $norm_{\rm cutoffpl}$ and QPO frequency are lower than the break frequency observed in the QPO fractional rms and QPO frequency relation.
However, the break frequency for the relation between $T_{\rm in}$ and QPO frequency occurs at a higher frequency.


\begin{table}
 \centering
 \caption{
    The best-fitting spectral parameters obtained using the model \texttt{constant * tbabs * (diskbb + relxill + cutoffpl)}. 
    }
 \label{tab:2}
    \setlength{\tabcolsep}{1.5pt} 
    \tablewidth{0.8\columnwidth}
    \tabletypesize{\small}
    \renewcommand{\arraystretch}{1.4}
    \begin{tabular}{c|cc|cccccc|c|c}
        \toprule 
        \toprule

   Component & \multicolumn{2}{c|}{\texttt{diskbb}}  & \multicolumn{6}{c|}{\texttt{relxill}} & \texttt{cutoffpl} &\multirow{2}{*}{$\chi^2/d.o.f $} \\
         \cline{1-10}
     
        
   ExpIDs$^{*}$ &  $ T_{\rm in}$[keV] & $norm$$[10^4] $  &  $R_{\rm in}$[$R_{\rm g}$] & $\Gamma$ & \(\log \xi\) & $E_{\text{cut}}$[keV] & $R_{\text{ref}}$ & $norm$  & $norm$ &   \\
  
        \cline{1-11}
  
  1101 & $0.440\pm0.013$ & $7.2_{-1.1}^{+1.2}$ &  $-1.5\pm0.2$ & $1.69_{-0.01}^{+0.02}$ & $3.71_{-0.05}^{+0.13}$& $28.9\pm0.7$  & $0.43_{-0.06}^{+0.09}$ & $0.29\pm0.02$ & $0.61\pm0.04$ 
        & $1475.1/1336$ \\
  
        1102 & $0.453\pm0.012$ & $6.5_{-0.8}^{+1.0}$ & $-1.5\pm0.2 $ & $1.69_{-0.03}^{+0.01}$ &  $3.73_{-0.06}^{+0.10} $  & $29.0_{-0.8}^{+0.9}$ & $0.45_{-0.08}^{+0.15}$ & $0.28_{-0.03}^{+0.02}$ & $0.57\pm0.04$
        & $1444.0/1336$ \\
  
        1104 & $0.445_{-0.012}^{+0.013}$ & $6.7\pm1.0$     &  $-1.49_{-0.16}^{+0.14}$ & $1.673_{-0.018}^{+0.019}$ & $3.72_{-0.04}^{+0.06} $ & $29.3\pm0.7$ & $0.51_{-0.07}^{+0.11}$ & $0.27\pm0.02$ & $0.60\pm0.03$
        & $1705.6/1335 $ \\
     
        1304 & $0.55\pm0.02$ & $2.1\pm0.3$ & $-1^{**}$  & $1.99\pm0.02$ & $3.7^{**}$ & $36\pm2$ & $0.28\pm0.04$ & $0.42\pm0.018$ & $0.43_{-0.04}^{+0.03}$
        & $1318.6/1255$ \\
        
        1405 & $0.58\pm0.02$ & $1.6_{-0.2}^{+0.3}$ & $-1^{**}$& $2.00\pm0.03$  &$3.7^{**}$& $36_{-2}^{+3}$ & $0.30_{-0.04}^{+0.05}$ & $0.41_{-0.02}^{+0.03}$ & $0.43_{-0.05}^{+0.04}$
        & $1243.6/1231$ \\
        
        1701 & $0.64\pm0.02$ & $1.11_{-0.13}^{+0.14}$  &$-1^{**}$ & $2.05\pm0.03$ & $3.7^{**}$& $41\pm3$ & $0.32_{-0.04}^{+0.05}$ & $0.40\pm0.02$ & $0.39_{-0.05}^{+0.04}$
        & $1240.1/1225$ \\
         
        1804 & $0.81\pm0.02$ & $0.42\pm0.07$  &$-1^{**}$& $2.19\pm0.03$ &$3.7^{**}$& $44_{-4}^{+5}$ & $0.24\pm0.04$ & $0.53\pm0.04$ & $0.31\pm0.06$
         & $1371.7/1301 $ \\      
       
        1807 &$1.00\pm0.02$  & $0.24_{-0.03}^{+0.04}$  &$-1^{**}$ & $2.26_{-0.03}^{+0.02}$ & $3.7^{**}$ & $47_{-4}^{+5}$ & $0.18_{-0.03}^{+0.04}$ & $0.66\pm0.04$ & $0.30\pm0.06$
        & $1421.0/1276$ \\

        2003 & $1.086\pm0.005$ & $0.57\pm0.015$ & $-1^{**}$ & $2.36_{-0.03}^{+0.02}$ & $3.7^{**}$ & $43\pm3$ & $0.06\pm0.03$ & $0.71\pm0.03$ & $0.38\pm0.03$
        & $1512.1/1307$ \\
       
        2104 & $0.998\pm0.014$ & $0.34\pm0.03$  & $-1^{**}$& $2.29\pm0.02$  & $3.7^{**}$& $42\pm3$ & $0.14\pm0.03$ & $0.65\pm0.04$ & $0.37_{-0.04}^{+0.03}$
        & $1375.3/1281$ \\     
        
        2107 & $0.940\pm0.015$ & $0.35_{-0.03}^{+0.04}$ & $-1^{**}$ & $2.27_{-0.03}^{+0.02}$ & $3.7^{**}$& $48_{-4}^{+5}$ & $0.16\pm0.03$ & $0.61_{-0.04}^{+0.03}$ & $0.32\pm0.05$
        & $1252.0/1272$ \\
        
        2301 & $0.773_{-0.010}^{+0.011}$ & $0.46\pm0.05$ &$-1^{**}$& $2.21\pm0.02$ &$3.7^{**}$& $54_{-4}^{+6}$ & $0.24_{-0.02}^{+0.03}$ & $0.47\pm0.02$ & $0.24\pm0.05$
        & $1330.1/1336$ \\
      
        2401 & $0.760_{-0.010}^{+0.011}$ & $0.50_{-0.06}^{+0.05}$ & $-1^{**}$ & $2.21\pm0.02$ & $3.7^{**}$& $54_{-5}^{+6}$ & $0.24\pm0.03$ & $0.45\pm0.02$ & $0.26\pm0.05$
        & $1296.7/1308$ \\

        2503 & $0.885_{-0.011}^{+0.012}$ & $0.35\pm0.03$ & $-1^{**}$ &$2.30\pm0.02$ &$3.7^{**}$& $58_{-5}^{+7}$ & $0.17\pm0.02$ & $0.56\pm0.03$ & $0.23_{-0.05}^{+0.04}$
        & $1408.3/1334$ \\

        2702 & $1.022\pm0.005$ & $0.58\pm0.02$ &  $-1^{**}$ & $2.43\pm0.02$ &$3.7^{**}$ & $59_{-7}^{+9}$ & $0.13\pm0.03$ & $0.59\pm0.03$ & $0.27\pm0.04$
        & $1640.2/1338$ \\
        
        2703 & $1.024\pm0.005$ & $0.57\pm0.02$ &  $-1^{**}$& $2.45\pm0.02$ & $3.7^{**}$ & $65_{-8}^{+11}$ & $0.13\pm0.03$ & $0.62\pm0.03$ & $0.25\pm0.04$
        & $1563.0/1338$ \\
 
        2803 & $0.952_{-0.009}^{+0.010}$ & $0.36\pm0.02$  & $-1^{**}$& $2.36\pm0.02$ &$3.7^{**}$& $63_{-7}^{+9}$ & $0.21\pm0.03$ & $0.55\pm0.03$ & $0.23_{-0.05}^{+0.04}$
        & $1525.7/1338$ \\
        
        2905 & $0.976\pm0.010$ & $0.30\pm0.02$ & $-1^{**}$ & $2.428\pm0.015$ &$3.7^{**}$& $124_{-10}^{+12}$ & $0.17\pm0.03$ & $0.61\pm0.02$ & $^{***}$ 
        & $1231.3/1223$ \\
           
  \bottomrule 
 \end{tabular}

\begin{tablenotes}
\item[1] $^{*}$   We use the last four digits of each ExpID for clarity.
\item[2] $^{**}$  Fixed
\item[3] $^{***}$ The additional hard component (\texttt{cutoffpl}) is not needed in the spectrum of this ExpID through an F-test.

\end{tablenotes}

\end{table}

\section{Discussion}

In this work, we have studied the evolution of the type-C QPOs during the flaring state of \target\ using the \hxmt\ observations. 
We conducted a detailed investigation into the evolution of the QPO fractional rms amplitude with its frequency across different energy bands. The relation observed in low-energy bands differs significantly from that in high-energy bands (see Figure \ref{FIG:rms-frequency}). Interestingly, a common break was identified in both the 2--4 keV and 50--10 keV bands, though this break was not detected in the other energy bands.
Further spectral analysis reveals significant changes in the main spectral parameters around the break. 
Below we discuss our main results.

The physical origin of type-C QPOs in BHTs remains a subject of ongoing debate. Some evidence suggests that type-C QPOs may arise from a geometric effect \citep[e.g.,][]{Motta2015,Ingram2016}. Potential scenarios include the Lense-Thirring (LT) precession of the hot inner flow \citep{Ingram2009inner_flow} or the base of the jet \citep{Ma2021NatAs...5...94M}.
The LT precession of the hot inner flow is predicated on the assumption of a truncated disk. Variations in the inner radius of the disk account for the observed shift in QPO frequency (\citealt{Ingram2009inner_flow}). 
We found that the inner disk radius, as measured from the reflection spectrum fitting, remains close to the ISCO during the flaring state of \target\ we analyzed.
In this case, the decrease in $norm_{\rm diskbb}$ observed when the QPO frequency is below $\sim$4 Hz could be attributed to a change in the spectral hardening factor \citep{ZhangW2022}. The realistic radius does not change significantly during this period.
The constant inner disk radius observed contradicts the LT precession of the hot inner flow model.
Therefore, we propose that the small-scale jet precession model, which is independent of the accretion disk radius, provides a more plausible explanation for the physical origin of the type-C QPOs in \target\ \citep{Ma2021NatAs...5...94M}. In this model, the frequency of the QPO is related to the size of the jet base.

Below 10 keV, the QPO fractional rms amplitude generally decreases with increasing QPO frequency. 
This trend can be easily understood since these energy bands are primarily composed of a soft disk component and a hard thermal Comptonization component. The increase in QPO fractional rms towards higher energies suggests that the QPO modulation is primarily generated in the Comptonization region \citep{Zhang2020,LiuHX2022type-BQPO,Mendez2022Nature}.
As the QPO frequency increases, the spectrum becomes softer, characterized by a stronger disk component and a weaker Comptonization component, leading to a reduction in the QPO fractional rms amplitude.

In the energy range of 10 to 50 keV, the QPO fractional rms remains relatively stable, a trend that has also been observed in other sources, such as Swift J1658.2--4242 \citep{Xiaogc2019...SwiftJ1658}. 
In these bands, the spectrum is predominantly dominated by the hard thermal Comptonization component, where the contribution from the disk component is negligible.
The nearly constant QPO fractional rms is likely due to the fact that only a single component contributes to the flux modulation.



In the 50--100 keV energy band, we report the first detection of a break in the relationship between QPO fractional rms and its frequency. Below $\sim$4 Hz, the QPO fractional rms decreases with QPO frequency, while above $\sim$4 Hz, the QPO fractional rms remains relatively constant.
In the energy spectrum, an additional hard component is observed beyond 50 keV, alongside the standard thermal Comptonization component \citep{Peng20241727hardtail}. This extra hard component likely originates from a hybrid thermal/nonthermal Comptonization process at the base of the jet \citep{Cangemi2021}.
\cite{Yangzx2024ApJ1727} identified a QPO rms excess in a similar energy band, where the QPO fractional rms above $\sim$50 keV is significantly higher than that observed at lower energies.
They also found that the magnitude of the QPO-rms excess is strongly correlated with the flux of the additional hard component, suggesting that the QPO rms excess is likely produced within this additional hard component.
Notably, the additional hard component is still seen in the spectra of the flaring state of \target. From our spectral-fitting results, the flux of this component decreases with QPO frequency below $\sim$4 Hz and then flattens (see the evolution of the normalization of this component in Figure \ref{FIG:specpara}), similar to the trend of the QPO fractional rms versus frequency in the 50--100 keV band. Therefore, the increase in QPO fractional rms towards lower energies is likely due to the contribution of the additional hard component to the flux modulation.

We also observed a similar break in the relation between QPO fractional rms and frequency in the 2--4 keV energy band. However, this break is not evident in the 4--6 keV and 6--8 keV bands.
In the 2--4 keV energy band, the spectrum exhibits a significant contribution from the disk component (see Figure \ref{FIG:spec fitting}).
\citet{Ma2023} investigated the energy-dependent properties of the type-C QPO in MAXI J1820+070 across a broad energy range spanning 0.2--200 keV. Their analysis revealed that the phase lags of the QPO at low energies are consistent with those observed at high energies when referenced to 2 keV. They proposed that the innermost region of the accretion disk undergoes precession in a manner analogous to the base of the jet.
The break observed in the 2--4 keV band is likely related to changes in the disk. Indeed, we found that both the inner disk temperature and the normalization of the disk component exhibit flatter variations when the QPO frequency exceeds 4 Hz, leading to a corresponding flattening of the QPO fractional rms with frequency.
However, it is important to note that the changes in the disk component are strongly correlated with variations in the full reflection and additional Comptonization components. This suggests that the geometry of the accretion flow experiences significant changes around the break.

\begin{flushleft} 
\vspace{2\baselineskip}
This work is supported by the National Key R\&D Program of China (2021YFA0718500), the National Natural Science Foundation of China under grants 12333007, 12403053, U1838203, 12203052.    This work made use of data and software from the Insight-HXMT mission, a project funded by the China National Space Administration (CNSA) and the Chinese Academy of Sciences(CAS).
\end{flushleft}

\bibliography{apj_ref_1727}{}
\bibliographystyle{aasjournal}



\end{document}